# Magnetic & Chemical Non-Uniformity in $Ga_{1-x}Mn_xAs$

## as Probed with Neutron & X-Ray Reflectivity

B. J. Kirby

Department of Physics and Astronomy, University of Missouri, Columbia, Missouri 65211, USA and

Manuel Lujan Jr. Neutron Scattering Center, Los Alamos National Laboratory, Los Alamos, New Mexico 87545, USA

J. A. Borchers

NIST Center for Neutron Research, National Institute of Standards and Technology, Gaithersburg, Maryland 20899, USA

J. J. Rhyne

Manuel Lujan Jr. Neutron Scattering Center, Los Alamos National Laboratory, Los Alamos, New Mexico 87545, USA

K. V. O'Donovan

NIST Center for Neutron Research, National Institute of Standards and Technology, Gaithersburg, Maryland 20899, USA and

Department of Physiology and Biophysics, University of California, Irvine, California 92697

S. G. E. te Velthuis

Materials Science Division, Argonne National Laboratory, Argonne, Illinois 60439, USA

S. Roy

Department of Physics, University of California at San Diego, La Jolla, CA 92093 USA

Cecilia Sanchez-Hanke

National Synchrotron Light Source, Brookhaven National Laboratory, Upton, New York 11973, USA

T. Wojtowicz

Department of Physics, University of Notre Dame, Notre Dame, Indiana 46556, USA and

Institute of Physics of the Polish Academy of Sciences, 02-688 Warsaw, Poland

X. Liu, W. L. Lim, M. Dobrowolska and J. K. Furdyna

Department of Physics, University of Notre Dame, Notre Dame, Indiana 46556, USA

**Abstract**

We have used complementary neutron and x-ray reflectivity techniques to examine the depth profiles of a series of as-grown and annealed $Ga_{1-x}Mn_xAs$ thin films. A magnetization gradient is observed for two as-grown films and originates from a non-uniformity of Mn at interstitial sites, and not from local variations in Mn at Ga sites. Furthermore, we see that the depth-dependent magnetization can vary drastically among as-grown $Ga_{1-x}Mn_xAs$ films despite being deposited under seemingly similar conditions. These results imply that the depth profile of interstitial Mn is dependent not only on annealing, but is also extremely sensitive to *initial* growth conditions. We observe that annealing improves the magnetization by producing a new surface layer that is rich in Mn and O, indicating that the interstitial Mn migrates to the surface. Finally, we expand upon our previous neutron reflectivity study of $Ga_{1-x}Mn_xAs$, by showing how the depth profile of the chemical composition at the surface and through the film thickness is directly responsible for the complex magnetization profiles observed in both as-grown and annealed films.

## I. Introduction

There has been recent widespread interest in the development of ferromagnetic dilute magnetic semiconductors (DMS). The material at the forefront of this effort has been $Ga_{1-x}Mn_xAs$ [1,2], in which ferromagnetism originates from coupling among spin 5/2 $Mn^{2+}$ ions at Ga sites ($Mn_{Ga}$) [3,4] that communicate their spin orientation among each other via self-generated holes [5].

Forcing Mn into Ga sites requires low-temperature molecular-beam epitaxial (MBE) growth, which also promotes creation of Mn interstitial defects ($Mn_I$). This is unfortunate, as $Mn_I$ are double donors that reduce the ferromagnetic transition temperature ($T_C$) by neutralizing holes needed to mediate the ferromagnetic exchange [6,7,8]. $Mn_I$ also reduce the magnetization ($M$), as calculations suggest they share an antiferromagnetic exchange interaction with neighboring $Mn_{Ga}$ [8]. Therefore, control of Mn site occupation is critical for fabrication of $Ga_{1-x}Mn_xAs$ films of high ferromagnetic quality. Such control can be partially achieved through post-growth annealing, which has been shown to greatly increase $T_C$ [9,10,11] and $M$ [12] by diffusing $Mn_I$ to the film surface [13,14,15,16,17]. While this technique has produced $Ga_{1-x}Mn_xAs$ films with $T_C$ impressive for a true DMS (> 150 K [10,11]), researchers have yet to produce truly ideal $Ga_{1-x}Mn_xAs$ samples. In fact, *room temperature $T_C$ for $Ga_{1-x}Mn_xAs$ is theoretically possible*, but will require further advances in growth condition control [18], motivating efforts to understand how to reliably grow uniform films with the highest possible $Mn_{Ga}/Mn_I$ ratio, and to explore the Mn diffusion process during annealing. Reflectometry is a natural tool with which to study these topics, as its depth-sensitivity allows us to examine uniformity of Mn site occupation, and the role of vertical impurity diffusion in $Ga_{1-x}Mn_xAs$ films. In this paper, we present reflectivity measurements of a series of $Ga_{1-x}Mn_xAs$ films that were grown and annealed one after the other under very similar conditions. PNR was used to obtain depth-dependent magnetic and chemical composition [19,20,21],Cu $k_\alpha$ x-ray reflectometry (XRR) was used to establish detailed depth profiles of the films' surface layers [22], and resonant XRR was used to identify the chemical composition of the films' surface layers [23].

## II. Reflectivity

A sample's specular reflectivity as a function of wavevector transfer $R(Q)$ is dependent upon that sample's depth-dependent scattering length density $\rho(z)$. In general, in-plane features are averaged over [20]. For x-rays, $\rho(z)$ is dependent on the atomic numbers of the constituent elements [22] – making XRR sensitive to a film's depth-dependent chemical composition. For neutrons, $\rho(z)$ has both a chemical (nuclear) component ($\rho_{Chem}$)dependent upon the characteristic scattering lengths $b$ of the constituent elements, and a magnetic component ($\rho_{Mag}$) proportional to the in-plane sample magnetization $M$. For the case of polarized neutrons, the two non spin-flip reflectivities are sensitive to the depth-dependent chemical composition, and $M(z)$ (primarily the component parallel to $H$), and the two spin-flip reflectivities are sensitive to the component of $M(z)$ perpendicular to $H$ [19,20,21].

For $Ga_{1-x}Mn_xAs$ on a GaAs substrate, XRR and PNR are particularly complementary. PNR is quite sensitive to Mn doping in this system ($b_{Ga} = 7.288$ fm$^{-1}$, $b_{Mn} = -3.73$ fm$^{-1}$, 310% difference), while Cu k$_\alpha$ XRR is virtually insensitive to it ($Z_{Ga} = 31$, $Z_{Mn} = 25$, 11% difference). Therefore for such a sample, PNR is sensitive the chemical and magnetic composition of the entire sample, while XRR (which gave access to much smaller length scales than PNR) yields high resolution chemical depth profiles of material residing on the $Ga_{1-x}Mn_xAs$ free surface.

Quantitative information was extracted by model fitting to find a $\rho(z)$ model that reproduces the data with the lowest possible value of $\chi^2$ [20,24,25,26]. Uncertainties for various fitting parameters were estimated by individually perturbing those parameters away from their best-fit values, and examining the resulting increase in $\chi^2$. Perturbations that increase $\chi^2$ by greater than 1 correspond to models that do not fit the data within one standard deviation, and are deemed to be unacceptable [24].

## III. Sample Preparation

Three separate $Ga_{1-x}Mn_xAs$ films were MBE grown on GaAs substrates held in place by indium on a molybdenum block [13,14]. The substrate temperature was controlled via a thermocouple located behind the Mo-block. Following growth, each sample was cleaved into pieces - one piece for annealing (at 270 °C, for about 1 hour, in a $N_2$ environment), and one piece left as-grown. The result was three sets of as-grown/annealed pairs (denoted Set A, Set B, and Set C). Primary characterization of the samples was done by using x-ray diffraction to establish $x$ [27], and anomalous Hall Effect to measure the apparent hole concentration $p$, and $T_C$ [28]. The macroscopic sample properties for each Set are shown in Table 1. Despite being fabricated under very similar conditions, the as-grown piece of Set C has significantly higher $T_C$ and $n$ than the Set A and B as-grown counterparts. For all three Sets, an increase in $T_C$ and $p$ are observed after annealing.

## IV. Experimental Results

Rotating anode XRR measurements were conducted with Cu $k_\alpha$ radiation at Los Alamos National Laboratory's Lujan Neutron Scattering Center. Figure 1 shows the XRR data and fits for the as-grown and annealed samples. The data and fits are multiplied by $Q^4$ to better visualize regions of high and low $Q$ in the same plot. For each of the sample sets, the annealing-dependent differences in the reflectivities are striking. Since XRR is virtually insensitive to the $Ga_{1-x}Mn_xAs$/GaAs interface, these data constitute immediate evidence that annealing significantly altered the film surface – even without fitting. The $\rho(z)$ models used to fit the data are inset in Fig. 1 [29]. For all three sets, we observe that an oxidation layer is present on the surface of as-grown films, and that annealing added approximately 20 Å of additional material to the surface. The $\rho(z)$ models are non-uniform and lack sharp interfaces, suggesting a mottled aggregation of surface material.

To further examine the characteristics of the surface material, the Set A pieces were examined using resonant XRR at Beamline X13A of Brookhaven National Laboratory's National Synchrotron Light Source. In this measurement, sample reflectivity was recorded as a function of incident x-ray energy, at a fixed reflection angle (i.e. a fixed wave-vector transfer) [30]. In this way, elements can be detected via peaks or dips in the reflectivity corresponding to electronic transitions for that element [23]. Figure 2 shows the reflectivity as a function of energy for the as-grown and annealed samples. Both the as-grown and annealed films have similar sharp features around 450 eV, likely originating from indium contamination from the sample holder in the growth chamber [31]. However, it is the *annealing-dependent* features that are most interesting. The annealed sample displays much more distinct features near the oxygen K1s (543.1 eV), manganese $L_3 2p_{3/2}$ (638.7 eV), and the manganese $L_2 2p_{1/2}$ (649.9 eV) electron binding energies than does the as-grown sample. Since this type of measurement is most sensitive to composition near the film surface, these results strongly suggest that annealing increased the concentration of Mn and O at the film surface.

To explore the resultant magnetic properties, PNR measurements were taken for each of the samples after cooling them below 20 K while in an in-plane field of $H = 6.6$ kOe (26.4 $\pi$ A·m$^{-1}$). The spin-flip reflectivities were negligibly small for all samples, meaning we could detect no in-plane component of $M$ perpendicular to $H$. Figure 3 compares the as-grown and annealed non spin-flip reflectivities and fits for sets A and C. Since the difference between the two spin states originates from the sample $M$, the insets of Fig. 3 show the spin-up and spin-down reflectivities and fits manifested as spin asymmetry (the difference in spin-up and spin-down divided by their sum).

First, consider Set A. The frequency of the annealed sample's oscillations in the reflectivity (Fig. 3 top panel) is larger than that of the as-grown, implying an increase in sample thickness upon annealing. At low-$Q$ the as-grown and annealed samples have reflectivities of similar intensity, while at high-$Q$ the annealed sample's reflectivity is consistently more intense than that of the as-grown. Fitting reveals that this difference is due to an increase in the surface $\rho_{Chem}$ for the annealed sample as compared to the as-grown. The amplitudes of the low-$Q$ spin asymmetry peaks (Fig. 3 top panel inset) are clearly larger for the annealed sample, due to a large increase in net $M$. Especially at low-$Q$, the as-grown sample's peaks are smeared (less clearly resolved), while the annealed sample's peaks are more sharply defined. This smearing is especially important, as fitting reveals it to be evidence of a pronounced gradient in $M$.

Set C is different, as the high-$Q$ intensity of reflectivity (Fig. 3 bottom panel) is fairly similar for the as-grown and annealed samples, showing that annealing has a different effect on the surface $\rho_{Chem}$ relative to Set A. The Set C spin asymmetries (Fig. 3 bottom panel inset) also differ from Set A, as the two lowest-$Q$ peaks do not smear together for either the as-grown and annealed samples, evidence that *neither* sample has a $M$ gradient. Due to the small surface area of the as-grown Set C piece (3/8 the size of the other samples discussed), the high-$Q$ data quality is not as good as that of the other samples. While this makes precise quantitative assessment of the film thickness and surface composition more difficult, the low-$Q$ statistics are more than adequate to show that there is no $M$ gradient like the one observed for the as-grown piece of Set A.

The $\rho(z)$ models that produce the best fit to the Set A (top panel) and Set C (bottom panel) PNR data are shown in Figure 3, separated into chemical and magnetic components. $M$ is plotted relative to a separate scale on the right-hand axis. The $Ga_{1-x}Mn_xAs$ layer is clearly delineated from the GaAs substrate in each model, and corresponds to a region of decreased $\rho_{Chem}$, and non-zero $M$.

For Set A the best-fit $\rho(z)$ model for the as-grown sample consists of four layers: 20 Å oxide layer, 324 Å top $Ga_{1-x}Mn_xAs$ sub-layer, 200 Å bottom $Ga_{1-x}Mn_xAs$ sub-layer, and GaAs substrate. The surface layer is at most 33 Å thick, has a $\rho_{Chem}$ similar to the $Ga_{1-x}Mn_xAs$ layer, and zero $M$ [32]. $\rho_{Chem}$ does not vary between the two $Ga_{1-x}Mn_xAs$ sub-layers ($2.79 \times 10^{-6} \pm 2 \times 10^{-8}$ Å$^{-2}$), but $M$ of the top $Ga_{1-x}Mn_xAs$ sub-layer is nearly double that of the bottom $Ga_{1-x}Mn_xAs$ sub-layer [33]. The sample's net $M$ is 23 emu·cm$^{-3}$. The Set A annealed piece model requires only three layers: 40 Å surface layer, 524 Å $Ga_{1-x-}Mn_xAs$, and GaAs substrate. The surface layer is *at least* 33 Å thick, its $\rho_{Chem} = 3.08 \times 10^{-6} \pm 8\times10^{-8}$ Å$^{-2}$, and its $M$ is no greater than 29 emu·cm$^{-3}$ – thicker and of drastically different composition than the surface layer on the as-grown sample. The $Ga_{1-x}Mn_xAs$ layer has a constant $M = 52 \pm 6$ emu·cm$^{-3}$ (significantly greater than the surface), and $\rho_{Chem} = 2.84\times10^{-6} \pm 3\times10^{-8}$ Å$^{-2}$ (significantly lower than the surface). While the surface $\rho_{Chem}$ happens to match that of the substrate, x-ray results (Fig. 1 and Fig. 2) rule out the possibility of pure GaAs at the surface. The surface layer thus results from oxidation.

PNR measurements of Set B are discussed in detail in Ref.14, and the results are quite similar to those of Set A, even though the Set B films are nearly twice as thick. For Set B, there is a pronounced $M$ gradient that is smoothed upon annealing, and annealing is observed to add a surface layer of drastically different chemical composition than that of $Ga_{1-x}Mn_xAs$.

The chemical and magnetic profiles for Set C are very different from those for Sets A and B. For Set C (bottom panel of Fig. 4), both the as-grown and annealed models consist of three layers (non-magnetic surface layer, $Ga_{1-x}Mn_xAs$ layer, and GaAs substrate). The as-grown model has a constant $Ga_{1-x}Mn_xAs$ layer with $M = 27 \pm 8$ emu·cm$^{-3}$. The model suggests that the surface $\rho_{Chem}$ is much lower than that of $Ga_{1-x}Mn_xAs$, but the fitting is not very sensitive to this value. The model for the annealed sample has an increased net $M$, but the same flat $M$ distribution, with constant $Ga_{1-x}Mn_xAs$ $M = 39 \pm 3$ emu·cm$^{-3}$.

$\rho_{Chem}$ of the annealed surface is not drastically different from that of the as-grown $Ga_{1-x}Mn_xAs$, and does not feature the surface spike evident for Sets A and B.

In summary, the x-ray and neutron analysis for Sets A, B and C reveal several significant features in the $M$ and chemical profiles that change upon annealing. For the as-grown films Sets A and B, $M$ is reduced from the bulk values and is depleted significantly near the $Ga_{1-x}Mn_xAs$ /GaAs interface. (Note that this gradient was reported previously for Set B [14].) This $M$ gradient, however, is not evident for Set C despite similarities in the growth conditions. $M$ increases for Sets A and B upon annealing and the $M$ gradient flattens. While annealing also improves the $M$ profile in Set C, the change is not as dramatic.

Corresponding changes in the chemical profiles for sets A and B upon annealing include the addition of a 20 – 50 A surface layer that is composed of Mn and O. This surface layer is magnetically dead and is also not apparent in the annealed sample in Set C, For all three samples, the chemical composition is flat for both annealed and as-grown samples through the entire film depth.

## V. Magnetization Gradient

One of the most striking differences between the Set A as-grown and annealed samples is that the $M$ profile is graded before annealing and flat afterwards. To determine if this feature originates from chemical variations in the film, it is important to determine if this feature is statistically robust and to establish the uncertainties associated with this feature. The $M$ gradient can be characterized by two parameters,

$$R_M = \text{Top Magnetic Sub-layer } M \text{ / Bottom Magnetic Sub-layer } M,$$

which describes the $M$ falloff of the gradient, and

$R_T$ = Bottom Magnetic Sub-layer thickness / Top Magnetic Sub-layer thickness,

which describes the spatial extent of the gradient.

For the as-grown piece of Set A, the best-fit $R_M$ = 1.81, and the best-fit $R_T$ = 0.62 (i.e. a pronounced $M$ gradient).  For the annealed sample, the best-fit $R_M$ = 1.00, and the best-fit $R_T$ = 0.00 (i.e. no $M$ gradient).  To assess the level of certainty in these values, $R_M$ and $R_T$ were individually perturbed away from their best-fit values (corresponding to $\chi^2_0$) and the resulting effect on $\chi^2$ was monitored.  The left-hand panel of Figure 5 shows $\chi^2 - \chi^2_0$ vs. $R_M$ with $R_T$ fixed at 0.62 and 0.00 for the as-grown and annealed samples respectively, and the right-hand panel shows $\chi^2 - \chi^2_0$ vs. $R_M$ with $R_T$ fixed at 1.81 and 1.00 for the as-grown and annealed samples respectively.  This demonstrates that the best-fit models strongly favor a pronounced $M$ gradient for the as-grown sample, and zero $M$ gradient for the annealed.  Each of the insets in Fig. 5 shows the most similar as-grown and annealed $M$ models corresponding to fits that reproduce the data with one standard deviation uncertainty  ($\chi^2 - \chi^2_0$ = 1).  Even with this large deviation, the annealed $\rho(z)$ profiles have smoother $M$ gradients than the as-grown.

A similar uncertainty analysis of $\rho_{Chem}$ indicates that there is less than a 0.6 % change in $\rho_{Chem}$ across the depth of the $Ga_{1-x}Mn_xAs$ layer in either of the Set A pieces.  If we assume constant density, this implies that the concentration of Mn at Ga sites $x$ in the as-grown (annealed) sample changes by less than 0.008 (0.013) across the entire $Ga_{1-x}Mn_xAs$ layer.  However, a gradient in the concentration of *other* Mn impurities (such as $Mn_I$) would have less effect on $\rho_{Chem}$, since such impurities do not displace a Ga atom.  Therefore, if the $Mn_I$ concentration in the as-grown (annealed) sample were changing by less than 0.02 (0.04) across the entire $Ga_{1-x}Mn_xAs$ layer, we would be insensitive to it.

## VI.  Discussion and Conclusions

As-grown $Ga_{1-x}Mn_xAs$ samples, exemplified by the 524 Å and 1035 Å films in Sets A and B respectively, and samples described in Ref. 15, frequently exhibit a large $M$ gradient through the depth of the film (i.e., $M$ approximately doubles from substrate to surface.)  While our previous study first identified this gradient [14], our new results indicate that it is a signature of a depth-dependent concentration of Mn at the interstitial sites.  Specifically, our reflectivity data reveal that the concentration of Mn at the substitution sites ($Mn_{Ga}$) is constant through the thickness of the film within less than 1% and thus is not responsible for the large $M$ gradient.  Since a typical $Mn_I$ concentration is about 0.02 [13] and is below the sensitivity of our PNR measurements, our results can only be explained by a corresponding doubling of the $Mn_I$ concentration throughout the film depth that is responsible for the observed $M$ gradient.  Since the site occupation of Mn depends strongly on growth thermodynamics [13], we conclude that the mobility of Mn atoms was changing during the growth – possibly as a result of strong non-equilibrium growth, or even from time-dependent variations in the substrate temperature.

Our PNR results for the as-grown piece of Set C, which has a higher $T_C$ and $p$ than Sets A and B, shows a flat $M$ profile, indicative of a more uniform distribution of $Mn_I$ through the film thickness.  Since the Set C samples were grown under very similar conditions, this result further demonstrates that a $M$ gradient is NOT an inherent property of all as-grown $Ga_{1-x}Mn_xAs$ films.   As an explanation, we note that the growth temperature was controlled by a thermocouple located behind the Mo-block sample holder for Sets A, B and C.  Due to small variations in thermal conductivity for each Mo block, there were likely slight, non-systematic variations in the actual *surface* temperature of the deposited films,  It is probable that these small changes in substrate surface temperature give rise to drastically different $Mn_I$ depth-profiles.  These results illustrate the pronounced sensitivity of the depth-dependent Mn site occupation (and thereby the depth-dependent $M$) to the initial growth temperature and/or other growth conditions.

While our previous study [14] demonstrated that annealing increases the net $M$ and flattens any $M$ gradient, our new results reveal that these improvements in the ferromagnetic quality of the $Ga_{1-x}Mn_xAs$ films are a direct consequence of changes in the $Mn_I$ depth profile as well as changes in the chemical composition and thickness of the surface layer. Combining structural information obtained from XRR, resonant XRR and PNR for the annealed sample in Set A, it is clear that annealing adds a rough non-magnetic surface layer that rich in Mn and O. We conclude that Mn has migrated from interstitial sites to the film surface, freeing a greater number of Mn at Ga sites to participate in the ferromagnetic exchange. While Set B behaves in a similar manner, set C is different. The as-grown set C $M$ profile is already flat, and annealing has a less pronounced effect on the composition of the film surface. These observations are completely consistent with the lower as-grown $Mn_I$ concentration for Set C (i.e., there are fewer $Mn_I$ to send to the surface), and again illustrate the extreme sensitivity of $Mn_I$ site occupation to subtle variations in growth conditions.

In this study, we have thus exploited the complementary, but distinct, sensitivities of XRR, resonant XRR and PNR in order to obtain a comprehensive profile of the chemical composition and magnetic structure in a series as-grown and annealed $Ga_{1-x}Mn_xAs$ films prepared in nominally identical conditions. Our analysis reveals that features in the $M$ depth profile are directly linked to the depth profile of the Mn residing at interstitial, rather than at gallium, sites in the lattice. While annealing can improve the ferromagnetic properties of these films by driving the $Mn_I$ to the surface, careful control of growth conditions may be sufficient to produce $Ga_{1-x}Mn_xAs$ films of high magnetic quality.

Work at Missouri and Notre Dame was supported by National Science Foundation Grant DMR-0138195, and the Missouri University Research Reactor. Work at Los Alamos, Argonne, and Brookhaven was supported by the Office of Basic Energy Science, U. S. Dept. of Energy. Special thanks go to Chuck

[1] H. Ohno, A. Shen, F. Matsukura, A. Oiwa, A. Endo, S. Katsumoto, and Y. Iye, *Appl. Phys. Lett.* **69,** 363 (1996).

[2] Maciej Sawicki, *Journal of Magnetism and Magnetic Materials*, **300**, 1 (2006).

[3] T. Dietl, H. Ohno, F. Matsukura, J. Cibert, D. Ferrand, *Science* **287**, 1019 (2000)

[4] Richard Bouzerar, Georges Bouzerar, and Timothy Ziman, *Phys. Rev. B* **73**, 024411 (2006).

[5] B. Beschoten, P. A. Crowell, I. Malajovich, D. D. Awschalom, F. Matsukura, A. Shen, and H. Ohno, *Phys. Rev. Lett.* **83**, 3073 (1999).

[6] T. Wojtowicz, W. L. Lim, X. Liu, Y. Sasaki, U. Bindley, M. Dobrowolska, J. K. Furdyna, K. M. Yu, and W. Walukiewicz, *J. Superconductivity* **16**, 41 (2003).

[7] Georges Bouzerar, Timothy Ziman, and Josef Kudrnovsky, *Phys. Rev. B* **72**, 125207 (2005).

[8] J. Blinowski and P. Kacman, *Phys. Rev. B* **67**, 121204(R) (2003).

[9] T. Hayashi, Y. Hashimoto, S. Katsumoto, and Y. Iye, *Appl. Phys. Lett.* **78**, 1691 (2001).

[10] K. W. Edmonds, P. Boguslawski, K. Y. Wang, R. P. Campion, S. N. Novikov, N. R. S. Farley, B. L. Gallagher, C. T. Foxon, M. Sawicki, T. Dietl, M. B. Nardelli, and J. Bernholc, *Phys. Rev. Lett.* **92**, 37201 (2004).

[11] K. C. Ku, S. J. Potashnik, R. F. Wang, S. H. Chun, P. Schiffer, N. Samarth, M. J. Seong, A. Mascarenhas, E. Johnston-Halperin, R. C. Myers, A. C. Gossard, and D. D. Awschalom, *Appl. Phys. Lett.* **82**, 2302 (2003).

[12] S. J. Potashnik, K. C. Ku, S. H. Chun, J. J. Berry, N. Samarth, and P. Schiffer, *Appl. Phys. Lett.* **79**, 1495 (2001).

[13] K. M. Yu, W. Walukiewicz, T. Wojtowicz, I. Kuryliszyn, X. Liu, Y. Sasaki, and J. K. Furdyna, *Phys. Rev. B* **65**, 201303(R) (2002).

[14] B. J. Kirby, J. A. Borchers, J. J. Rhyne, S. G. E. te Velthuis, A. Hoffmann, K. V. O'Donovan, T. Wojtowicz, X. Liu, W. L. Lim, and J. K. Furdyna, *Phys. Rev. B* **69**, 81307(R) (2004).

[15] B. J. Kirby, J. A. Borchers, J. J. Rhyne, K. V. O'Donovan, T. Wojtowicz, X. Liu, Z. Ge, S. Shen, and J. K. Furdyna, *Appl. Phys. Lett.* **86**, 072506 (2005).

[16] D. Chiba, K. Takamura, F. Matsukura, and H. Ohno, *Applied Physics Letters* **82**, 3020 (2003).

[17] M. B. Stone, K. C. Ku, S. J. Potashnik, B. L. Sheu, N. Samarth, and P. Schiffer, *Appl. Phys. Lett.* **83**, 4568 (2003).

[18] T. Jungwirth, K. Y. Wang, J. Mašek, K. W. Edmonds, Jürgen König, Jairo Sinova, M. Polini, N.A. Goncharuk, A.H. MacDonald, M. Sawicki, A.W. Rushforth, R.P. Campion, L.X. Zhao, C.T. Foxon, and B.L. Gallagher, *Physical Review B* ,**72**, 165204 (2005).

[19] G. P. Felcher, *Phys. Rev. B* **24**, R1595 (1981).

[20] C. F. Majkrzak and M. R. Fitzsimmons (2005); Application of polarized neutron reflectometry to studies of artificially structured magnetic materials. In *Modern Techniques for Characterizing Magnetic Materials,* Zimei Zhu, editor. Kluwer.

[21] C.F. Majrkzak, K.V. O'Donovan, N.F. Berk (2006); Polarized neutron reflectometry. In *Neutron Scattering from Magnetic Materials,* T. Chatterji, editor. Elsevier.

[22] T. P. Russell, *Annual Review of Materials Science*, **21**, 249 (1991).

[23] *Center for X-Ray Optics and Advanced Light Source X-Ray Data Booklet*, Albert C. Thompson and Douglas Vaughan editors, (2nd edition, Lawrence Berkeley National Laboratory 2001).

[24] W.H. Press et al., *Numerical Recipes, the art of scientific computing*, (Cambridge University Press, Cambridge 1986) p. 534.

[25] P.A. Kienzle, K.V. O'Donovan, J.F. Ankner, N.F. Berk, C.F. Majkrzak; http://www.ncnr.nist.gov/reflpak. 2000-2006.

[26] J. F. Ankner and C. F. Majkrzak, *Neutron Optical Devices and Applications, SPIE Conference Proceedings,* Vol. 1738, edited by C. F. Majkrzak, and J. W. Wood (SPIE, Bellingham, WA, 1992), pp. 260-269.

[27] J. Sadowski, R. Mathieu, P. Svedlindh, J. Z. Domagala, J. Bak-Misiuk, K. Swiatek, M. Karlsteen, J. Danski, L. Ilver, H. Asklund, and U. Sodervall, *Appl. Phys. Lett.* **78**, 3271 (2001).

[28] $T_C$ was established via an Arrott Plot, in which the square of the Hall magnetization is plotted vs. magnetic field.

[29] For clarity in comparison, the three models are normalized such that identical $\rho(z)$ features occur at the same value of $z$.

[30] Attempts to measure the wave-vector dependent reflectivities at the Mn edges were unsuccessful, likely due to the low Mn concentration and absorption by the surface material.

[31] The presence of small amounts of indium is inconsequential. We have grown and annealed $Ga_{1-x}Mn_xAs$ samples with and without indium in the chamber, and have found it to have no effect on the resulting samples.

[32] There is very little certainty in $M$ for the surface layer, but the best fit requires reduced surface $M$.

[33] The functional form of the magnetic roughness between the two sub-layers is a Gaussian with a full-width at half maximum equal to the thickness of the lower sub-layer. The fits are not very sensitive to the functional form, but clearly favor a rough magnetic interface between the two sub-layers over a sharp one.

| Set | Ga$_{1-x}$Mn$_x$As layer thickness (Å) | $x$, Mn$_{Ga}$ concentration | As-Grown $p$ (290 K) ($10^{19}$cm$^{-3}$) | Annealed $p$ (290 K) ($10^{19}$cm$^{-3}$) | As-Grown $T_C$ (K) | Annealed $T_C$ (K) | As-grown saturation net $M$ (emu·cm$^{-3}$) | Annealed saturation net $M$ (emu·cm$^{-3}$) |
|---|---|---|---|---|---|---|---|---|
| A | 524 | 0.092 | 5.97 | 12.1 | 60 | 120 | 23 | 52 |
| B | 1035 | 0.076 | 5.08 | 11.1 | 60 | 125 | 17 | 48 |
| C | 530 | 0.081 | 9.78 | 21.2 | 70 | 130 | 27 | 39 |

**Table 1:  Summary of the Ga$_{1-x}$Mn$_x$As layer thicknesses, Mn$_{Ga}$ concentrations, hole concentrations, Curie temperatures, and saturation magnetizations of the three as-grown/annealed pairs discussed in this paper.**

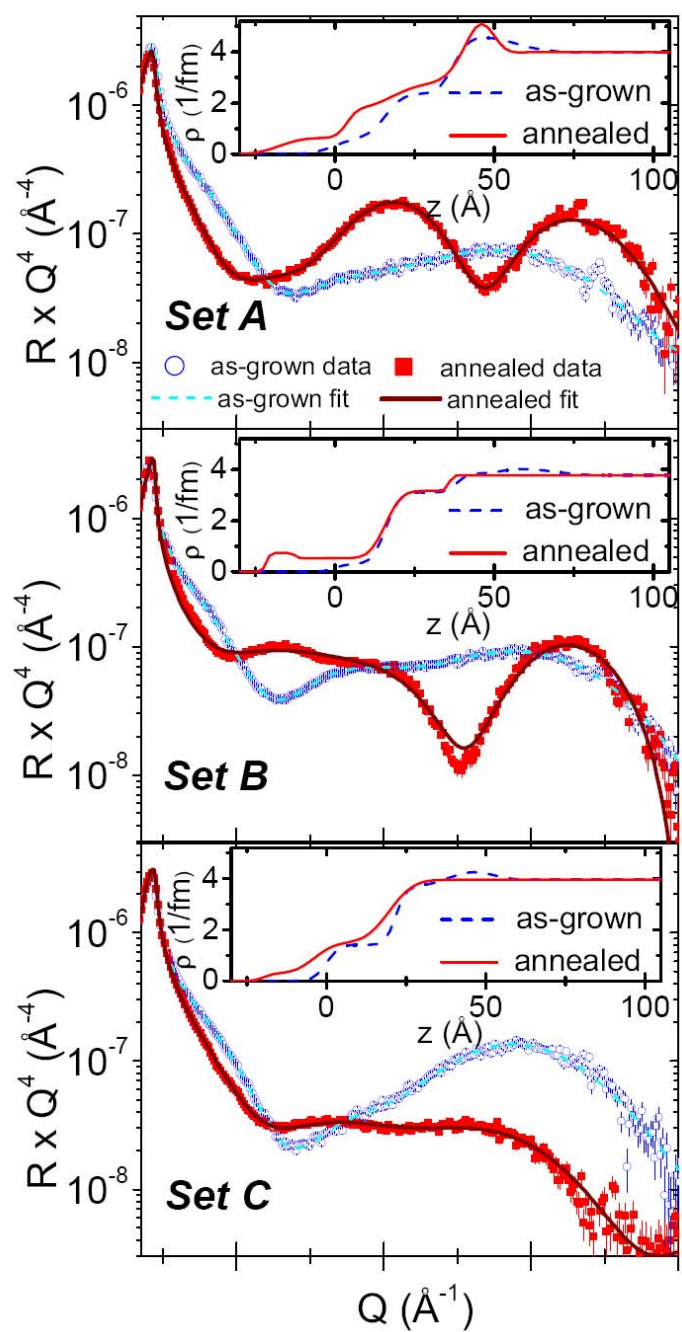

**Figure 1 (Color online):** XRR data, and fits for the as-grown and annealed films. The scattering length density depth profiles $\rho(z)$ used to fit the data are shown in the insets.

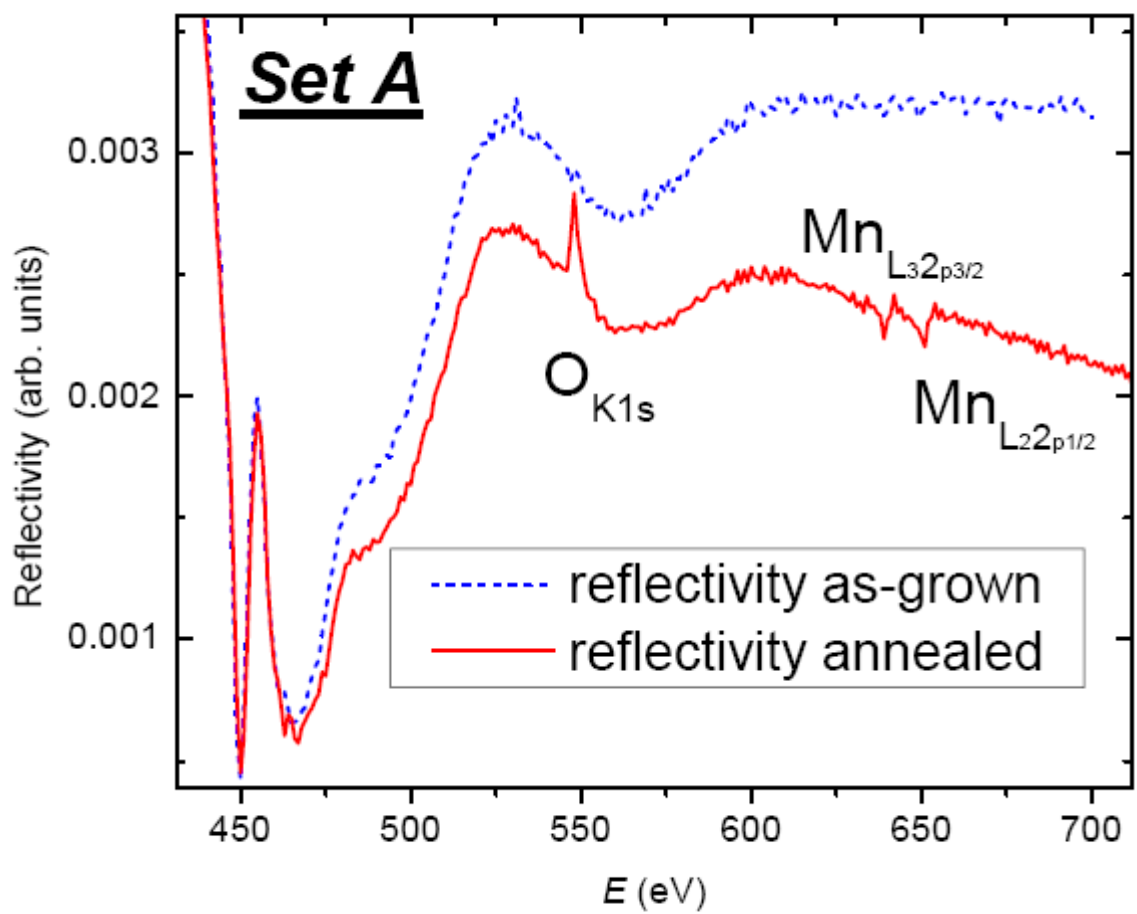

**Figure 2 (Color online): Resonant XRR data for the as-grown and annealed set A films. The annealed film features pronounced O and Mn peaks while the as-grown does not.**

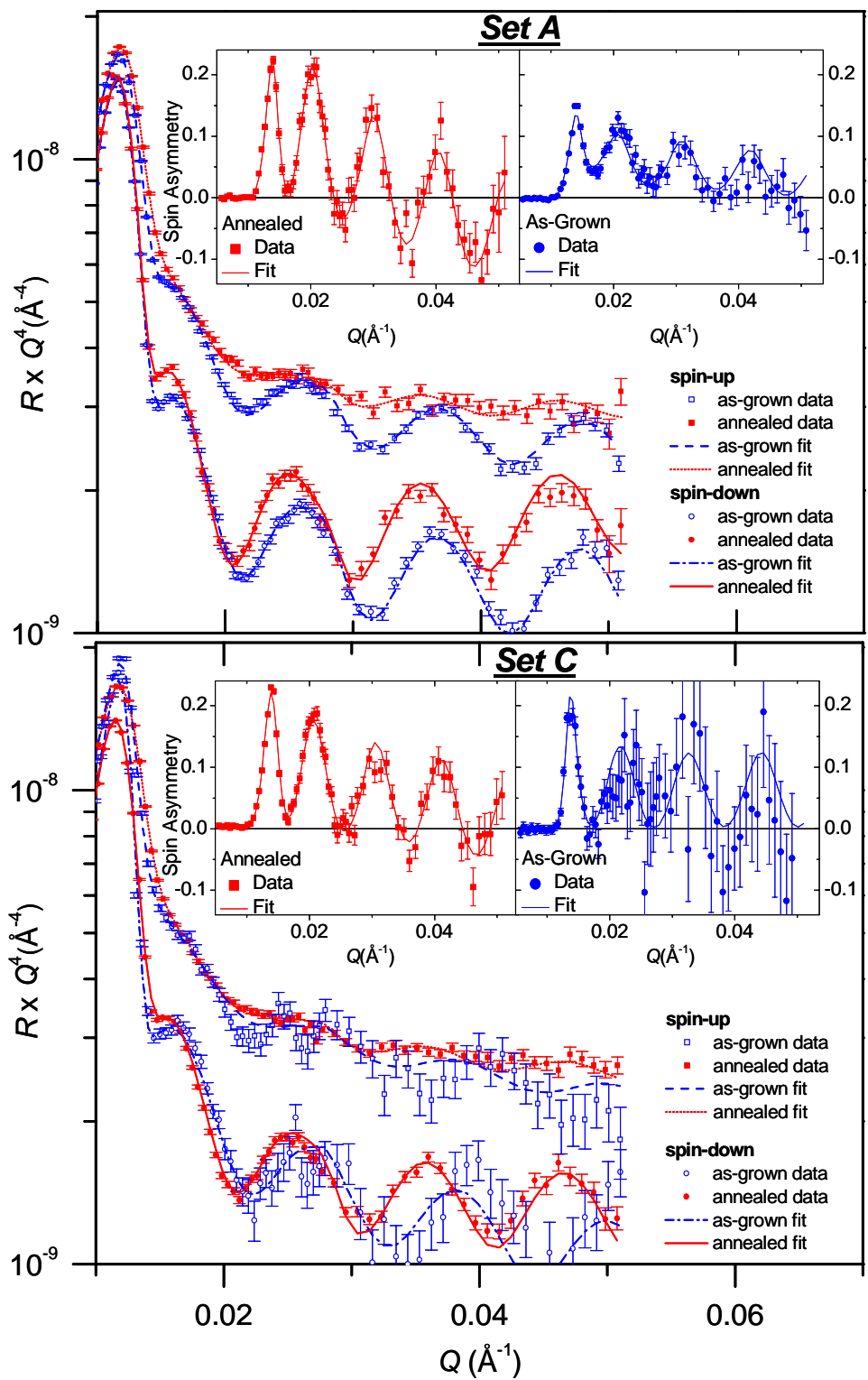

**Figure 3 (Color online): Spin-up and Spin-down neutron reflectivities and fits for the as-grown and annealed Set A and Set C samples. The spin-up data and fits are shown offset by $6\times10^{-10}$ Å$^{-4}$. The data and fits recast as spin asymmetry is shown in the inset.**

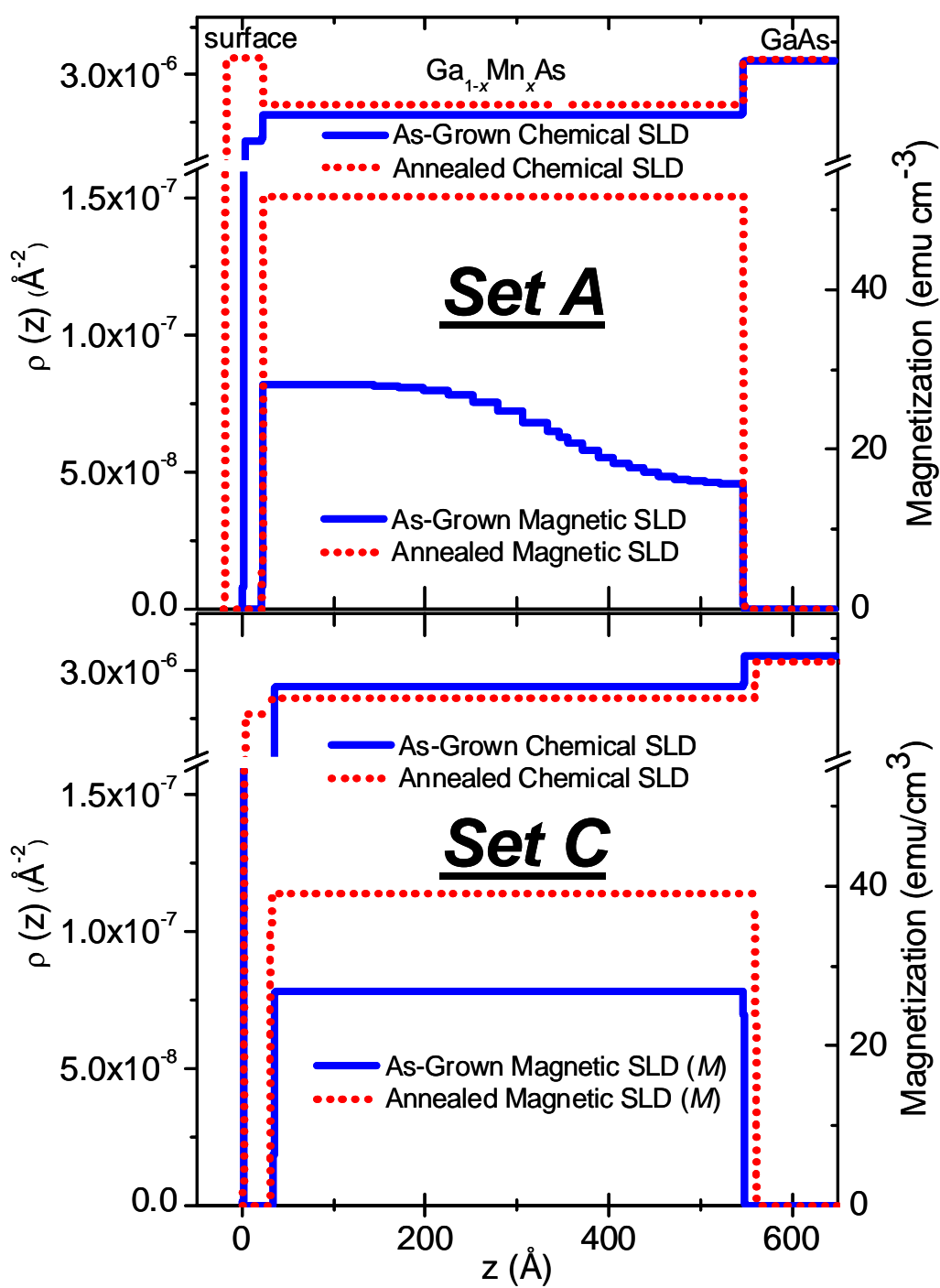

**Figure 4 (Color online):** Scattering length density depth profiles $\rho(z)$ used to fit the Set A and Set C data in Fig. 3.

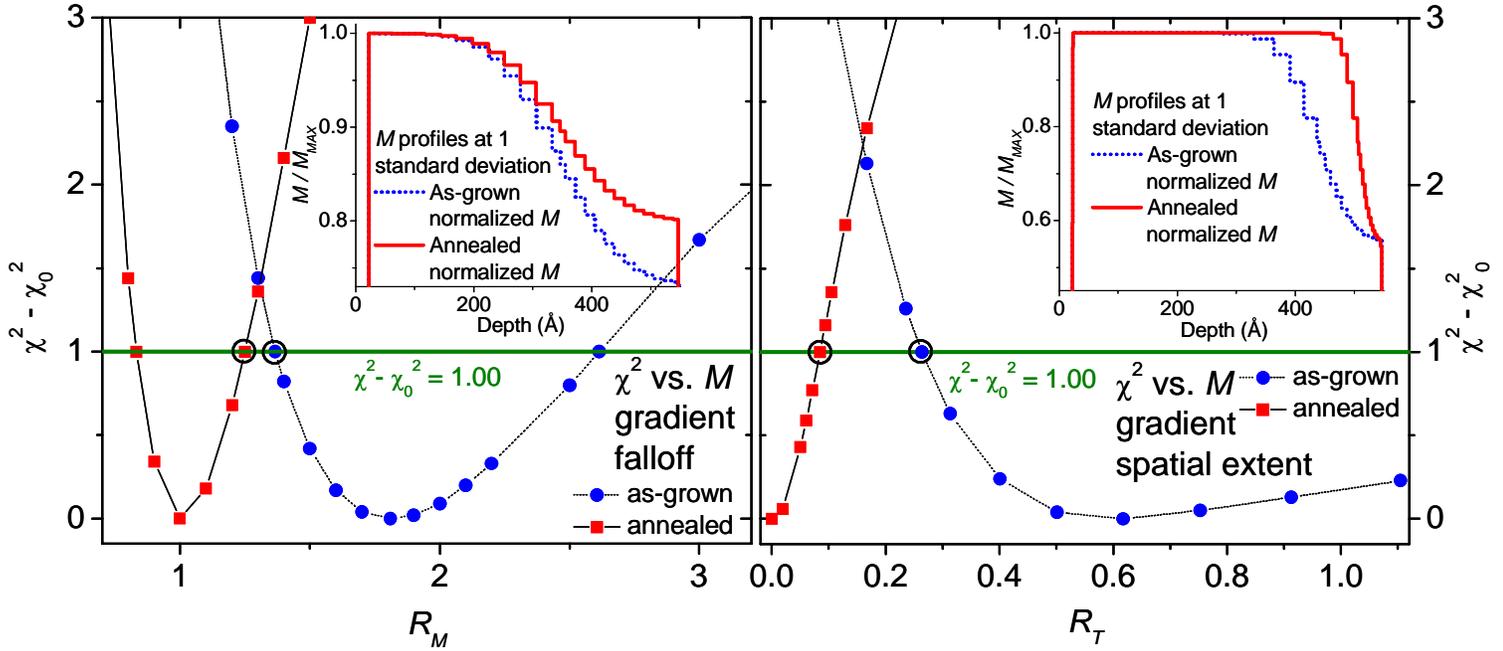

**Figure 5 (Color online):** Change in best-fit $\chi^2$ as a function of $M$ gradient falloff $R_M$ (left), and spatial extent $R_T$ (right), for the Set A as-grown and annealed samples. The inset $M$ models correspond to fits that reproduce the data with one standard deviation uncertainty (circled data points). $M$ is normalized by the maximum value of $M$ for each sample to allow for direct comparison. The inset models still show a clear smoothing of $M$ after annealing.

Figure Captions

1. **Figure 1 (Color online): XRR data, and fits for the as-grown and annealed films. The scattering length density depth profiles $\rho(z)$ used to fit the data are shown in the insets.**

2. **Figure 2 (Color online): Resonant XRR data for the as-grown and annealed set A films. The annealed film features pronounced O and Mn peaks while the as-grown does not.**

3. **Figure 3 (Color online): Spin-up and Spin-down neutron reflectivities and fits for the as-grown and annealed Set A and Set C samples. The spin-up data and fits are shown offset by $6\times10^{-10}$ Å$^{-4}$. The data and fits recast as spin asymmetry is shown in the inset.**

4. **Figure 4 (Color online): Scattering length density depth profiles $\rho(z)$ used to fit the Set A and Set C data in Fig. 3.**

5. **Figure 5 (Color online): Change in best-fit $\chi^2$ as a function of $M$ gradient falloff $R_M$ (left), and spatial extent $R_T$ (right), for the Set A as-grown and annealed samples. The inset $M$ models correspond to fits that reproduce the data with one standard deviation uncertainty (circled data points). $M$ is normalized by the maximum value of $M$ for each sample to allow for direct comparison. The inset models still show a clear smoothing of $M$ after annealing.**